\documentclass[aip,jcp]{revtex4-1}

\usepackage{graphicx}
\usepackage{dcolumn}
\draft 

\begin{document}

\title{Challenges for Machine Learning Force Fields in Reproducing Potential Energy Surfaces of Flexible Molecules} 

\author{Valentin Vassilev-Galindo}
\author{Gregory Fonseca}
\author{Igor Poltavsky}
\author{Alexandre Tkatchenko}
\email{alexandre.tkatchenko@uni.lu}
\affiliation{Department of Physics and Materials Science, University of Luxembourg, L-1511 Luxembourg City, Luxembourg}

\date{\today}

\begin{abstract}
\noindent 
Dynamics of flexible molecules are often determined by an interplay between local chemical bond fluctuations and conformational changes driven by long-range electrostatics and van der Waals interactions. This interplay between interactions yields complex potential-energy surfaces (PES) with multiple minima and transition paths between them. In this work, we assess the performance of state-of-the-art Machine Learning (ML) models, namely sGDML, SchNet, GAP/SOAP, and BPNN for reproducing such PES, while using limited amounts of reference data. As a benchmark, we use the {\itshape cis} to {\itshape trans} thermal relaxation in an azobenzene molecule, where at least three different transition mechanisms should be considered. Although GAP/SOAP, SchNet, and sGDML models can globally achieve chemical accuracy of 1~kcal mol\textsuperscript{-1} with fewer than 1000 training points, predictions greatly depend on the ML method used as well as the local region of the PES being sampled. Within a given ML method, large differences can be found between predictions of close-to-equilibrium and transition regions, as well as for different transition mechanisms. We identify key challenges that the ML models face in learning long-range interactions and the intrinsic limitations of commonly used atom-based descriptors. All in all, our results suggest switching from learning the entire PES within a single model to using multiple local models with optimized descriptors, training sets, and architectures for different parts of complex PES.
\end{abstract}

\maketitle

\section{Introduction}\label{sec:introduction}
Thermodynamic and dynamical properties of molecules can be computed if an accurate model for a potential-energy surface (PES) is provided. Among these properties, transition paths connecting pairs of minima on the PES are crucial for understanding the dynamics of complex systems,\cite{Hanggi1990} such as conformational changes in molecules,\cite{Bachmann2005a,Chakraborty2019} nucleation events during phase transitions,\cite{Cook2019,Zhang2019} folding and unfolding of proteins.\cite{Dobson2003,Piana2012,Stiller2019} 
The state-of-the-art methods for finding transition pathways range from the optimization of a single direction on the PES\cite{Peng1993,Peng1996} or a chain of states connecting both minima, e.g. the string\cite{E2007} and nudged elastic band (NEB)\cite{Jonsson1998,Henkelman2000} methods, to the more sophisticated transition path sampling techniques.\cite{Dellago1998,Bolhuis2018} Most of them often provide only a single ``optimal" transition path. The rate of success to find the path highly depends on the dimensionality and complexity of the PES: flexible molecules containing a few tens of atoms, such as organic photoswitches and peptides, are already challenging to deal with. Moreover, due to the non-trivial interplay between covalent and non-covalent interactions, the transitions in such molecules may happen following several different pathways. In this case, one needs to consider the contribution of every path to the transition process, and the knowledge about just one optimal pathway is insufficient. Practical studies of such transitions require reliable force fields (FF) able to accurately reconstruct broad regions of the PES, including multiple local minima and all the relevant pathways connecting them.

In recent years, the use of Machine Learning (ML) in chemistry and materials science has been a subject of intensive research.\cite{Chmiela2017,Chmiela2018,Sauceda2019,Chmiela2019c,Botu2015,Christensen2019,Christensen2020,Rupp2015,Glielmo2017,Eickenberg2018,Bartok2010,Bartok2015,Bartok2013,Li2015,Podryabinkin2017,Dral2017,Noe2019,Mardt2018,Behler2007b,Behler2011a,Behler2011,Behler2007,Jose2012,Behler2016,Gastegger2017,Schutt2018,Schutt2017a,Schutt2019,Rupp2012,Hansen2013,De2016,Artrith2017,Bartok2017,Yao2017,Faber2017,Glielmo2018,Grisafi2018,Tang2018,Pronobis2018,Faber2018,Ryczko2018,Zhang2018,Shao2016,Yao2018,Brockherde2017,Smith2018,Hansen2015} 
Namely, the advent of ML potentials has offered new tools for meeting the constantly increasing demand for accurate simulations of realistic systems, since such potentials aim to keep the accuracy of {\itshape ab initio} calculations with an efficiency closer to that of classical force fields. Among the available methodologies, neural networks (NN)\cite{Behler2007,Behler2011,Jose2012,Behler2016,Gastegger2017,Schutt2018,Behler2007b,Behler2011a,Mardt2018} and kernel-based methods\cite{Chmiela2017,Chmiela2018,Sauceda2019,Chmiela2019c,Bartok2010,Bartok2013,Bartok2015,Botu2015,Rupp2015,Li2015,Eickenberg2018,Podryabinkin2017,Glielmo2017,Christensen2019,Christensen2020} are the most used to learn the PES of molecules. However, this learning task is not easy and it has encouraged the improvement of data sampling,\cite{Li2015,Podryabinkin2017,Mardt2018,Dral2017,Noe2019} molecular representations\cite{Bartok2010,Bartok2015,Rupp2015,Eickenberg2018,Glielmo2017,Rupp2012,Hansen2013,De2016,Artrith2017,Bartok2017,Yao2017,Faber2017,Glielmo2018,Grisafi2018,Tang2018,Pronobis2018,Faber2018} and NN architectures.\cite{Behler2007,Jose2012,Behler2016,Gastegger2017,Schutt2018,Ryczko2018,Zhang2018} 
Also, some efforts have been directed towards improving ML-aided search and sampling of transition states and pathways.\cite{Noe2019,Meyer2019,Pattanaik2020} 
For instance, Noe \textit{et al.}\cite{Noe2019} showed a promising method to sample rare events between equilibrium states using Boltzmann generators. The method is by many orders of magnitude more efficient than “brute force” molecular dynamics (MD) simulations. Other approaches\cite{Meyer2019,Pattanaik2020} are built on  state-of-the-art methods for calculating transition states enhanced with ML techniques. ML-enhanced transition state search methods are more efficient than their precursors but present the same limitations. ML methods are often data demanding, making their application infeasible when computationally expensive \textit{ab initio} methods are required. Hence, constructing robust ML models for flexible molecules is the necessary next step for practical applications of ML potentials in chemistry and biology.  

There are two main challenges for building accurate ML models for flexible molecules. First, generating enough data around the transition regions of the PES. Second, building a highly accurate and data-efficient ML model that describes the resulting complex PES. In this work, we address both of these challenges on the example of an azobenzene (C$_{12}$H$_{10}$N$_2$) molecule. While being small in size, azobenzene is flexible enough to feature a {\itshape cis} to {\itshape trans} thermal relaxation following at least three possible channels: a rotation, an inversion and a rotation assisted by inversion mechanisms.\cite{Cattaneo1999,Cembran2004,Gagliardi2004, Wang2007, Tavadze2018} We start by discussing the problem of building reliable reference datasets for these transitions. Then, we assess the performance of state-of-the-art ML methods on the prediction of forces and energies along the obtained transition paths. The methods include NNs, on the example of Behler-Parrinello neural networks (BPNN)\cite{Behler2007,Behler2011} and SchNet\cite{Schutt2017a,Schutt2018,Schutt2019} architectures, and kernel-based methods, on the example of sGDML\cite{Chmiela2017,Chmiela2018,Chmiela2019c,Sauceda2019} and Gaussian Approximation Potentials (GAP)\cite{Bartok2010,Bartok2015} using the Smooth Overlap of Atomic Positions (SOAP) representation.\cite{Bartok2013} To highlight how the complexity of learning the PES increases with the flexibility of a molecule, we compare the results for azobenzene with those for a simpler glycine molecule. We limit the training datasets to 1000 geometries. The ML models unable to predict the PES of the considered small molecules correctly within this limit would face considerable problems for large flexible molecules where the cost of reference calculations increases very steeply.

The structure of this manuscript is the following: in section~\ref{sec:section_II} we present the isomers of glycine and azobenzene, build possible transition paths between them and construct the datasets for training ML models. In section~\ref{sec:section_III} we discuss the frequent pitfalls of state-of-the-art ML methods that describe different configurations of flexible molecules. Then in section~\ref{sec:section_IV}, we assess the performance of ML models (on the example of sGDML,\cite{Chmiela2017,Chmiela2018,Chmiela2019c,Sauceda2019} GAP/SOAP,\cite{Bartok2010,Bartok2015,Bartok2013} BPNN\cite{Behler2007,Behler2011} and SchNet\cite{Schutt2017a,Schutt2018,Schutt2019}) for azobenzene and glycine molecules trained for both equilibrium states, as well as the transition paths between them. In section~\ref{sec:section_V}, we describe the challenges of ML force fields when applied to flexible molecules. Section~\ref{sec:conclusions} contains the conclusions and an outlook.

\section{Constructing Reference Datasets for Isomerization}\label{sec:section_II}
The starting point for building any ML force field (MLFF) is collecting reference data covering the relevant parts of the PES of interest. When modeling transition pathways, the reference data can be split into two parts: (i) data covering the vicinity of the equilibrium states between which the transition process happens and (ii) data of `far-from-equilibrium' parts of the PES defining the transition path(s). While the equilibrium states are normally readily available, configurations describing the transition paths connecting them are, in most cases, not trivial to find. Moreover, the complexity of this task rapidly grows with the increase in flexibility and size of the molecule. In view of this, we had to employ two different strategies for generating the datasets for glycine and azobenzene isomerization. Below, we discuss in detail the process followed for each molecule separately. All calculations (unless specified otherwise) were performed in FHI-aims software\cite {Blum2009} using the Perdew-Burke-Ernzerhof (PBE) exchange-correlation functional\cite{Perdew1996} with tight settings and the Tkatchenko-Scheffler (TS) method\cite{Tkatchenko2009} to account for van der Waals (vdW) interactions. For all MD simulations, the i-PI package\cite{Kapil2018} was wrapped with FHI-aims. Detailed information of all methods and MD simulations, as well as all relevant configurations and datasets, are available in the supporting material.

\subsection{Glycine}

Glycine, being a rather small molecule, possesses numerous planar and non-planar conformers in the gas phase whose relative energies have been extensively studied.\cite{Csaszar1992,Pacios2001} Here, we consider the isomerization from the global minimum geometry, called Ip, to the IIIp conformer because it is the closest ``directly connected'' minimum. Transitions to any other metastable state from Ip go through this conformer. The Ip -- IIIp transformation can be characterized by a change of torsional angles $\tau$\textsubscript{1} and $\tau$\textsubscript{2} (see Fig.~\ref{fig:Fig1}), both around the C bond. They go from 180.0$^{\circ}$ and 0.0$^{\circ}$ in the isomer Ip to 0.0$^{\circ}$ and 180.0$^{\circ}$ in the isomer IIIp, respectively.

\subsubsection{Transition Path.}
To construct the transition path between the equilibrium states of glycine we used the string\cite {E2007} and NEB\cite {Jonsson1998,Henkelman2000} methods. Both methods converge without any issues providing similar pathways. The transition state obtained by employing the PBE+TS method lie only 2.4 kcal mol\textsuperscript{-1} above the Ip isomer (see Fig. S1 in the supporting material). The mechanism is defined by almost equal rotations of both $\tau$\textsubscript{1} and $\tau$\textsubscript{2} torsional angles (see table S1 in the supporting material for important geometric details).

\subsubsection{Dataset.} 
Since the relative energy between the Ip isomer and the highest-energy structure found on the minimum energy path (MEP) is less than 3 kcal mol\textsuperscript{-1}, the Ip -- IIIp transformation is accessible via standard constant-temperature MD simulations. So, to construct the dataset for glycine isomerization, we ran two dynamics starting from both equilibrium geometries. A total of 5000 configurations at 500 K with a timestep of 1 fs were obtained from each simulation. A transition was observed in the simulation starting from the IIIp isomer. 

\begin{figure}
\includegraphics[width=0.75\textwidth]{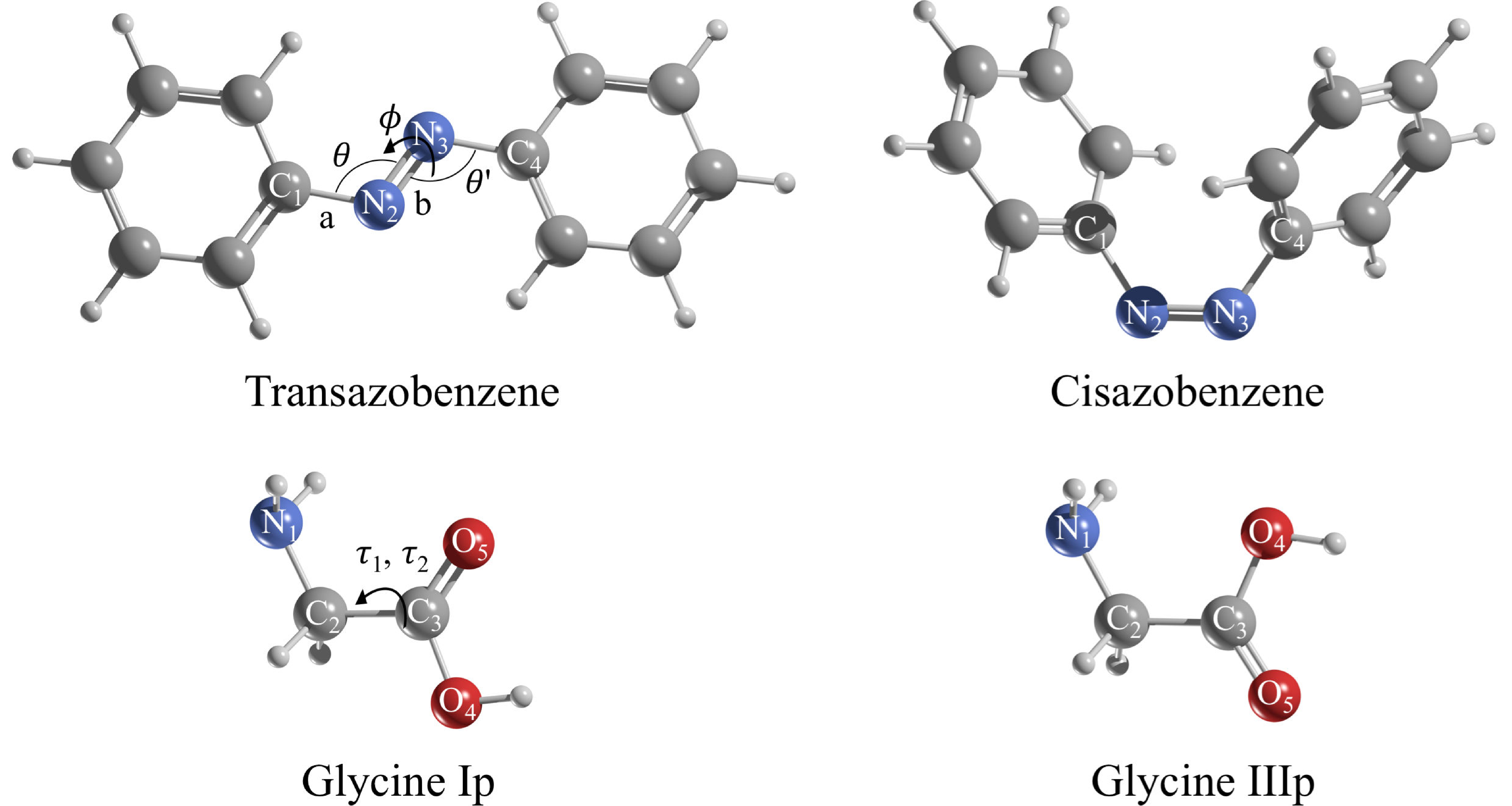}
\caption{\label{fig:Fig1} Optimized configurations of the minima considered in this work and labels of the main degrees of freedom: the bonds {\itshape a} (C\textsubscript{1}-N\textsubscript{2} and N\textsubscript{3}-C\textsubscript{4} in azobenzene) and {\itshape b} (N\textsubscript{2}=N\textsubscript{3} in azobenzene); the bending angles $\theta$ and $\theta$$'$ (C\textsubscript{1}-N\textsubscript{2}=N\textsubscript{3} and N\textsubscript{2}=N\textsubscript{3}-C\textsubscript{4} in azobenzene); and the torsional angles $\phi$ (C\textsubscript{1}-N\textsubscript{2}=N\textsubscript{3}-C\textsubscript{4} in azobenzene), $\tau$\textsubscript{1} and $\tau$\textsubscript{2} (N\textsubscript{1}-C\textsubscript{2}-C\textsubscript{3}-O\textsubscript{4} and N\textsubscript{1}-C\textsubscript{2}-C\textsubscript{3}=O\textsubscript{5} in glycine).}
\end{figure}

\subsection{Azobenzene}

Azobenzene is a photochemical compound, however it also exhibits a {\itshape cis} to {\itshape trans} thermal relaxation, on which we focus in this work. Hence, we avoid issues with electronic multi-reference states in azobenzene and use the semi-local DFT-PBE functional and include vdW interactions with the TS-vdW method for generating the reference data. The {\itshape cis} and {\itshape trans} configurations of azobenzene (Fig.~\ref{fig:Fig1}) differ mostly by a change in torsional angle $\phi$ around the N=N double bond from close to 10.0$^{\circ}$ to 180.0$^{\circ}$ during the isomerization. Although the existence of the two forms has been known since the works of Hartley in the 1930's,\cite{Hartley1937,Hartley1938} there is still an open debate regarding whether azobenzene primarily follows a rotation (changes around the dihedral angle $\phi$), an inversion (changes of the angles $\theta$ or $\theta$$'$) or a rotation assisted by inversion (changes of both $\phi$, and $\theta$ or $\theta$$'$) mechanism. DFT, Multi-reference methods and ML approaches have been used in an attempt to unveil the actual mechanism of isomerization,\cite{Cembran2004,Gagliardi2004,Diau2004,Bandara2012,Wang2007,Tavadze2018,Cattaneo1999,Dokic2009} but conclusive evidence favouring a particular mechanism has yet to be found. 

\subsubsection{Transition Paths.} 
Although azobenzene is not much larger than glycine, all the transition pathways which can be found in the literature for this molecule are constructed manually.
One can easily check that neither the string nor NEB methods converge to a reasonable path for {\itshape cis} to {\itshape trans} transition. 
Following the previous works~\cite{Cattaneo1999,Cembran2004,Gagliardi2004, Wang2007, Tavadze2018} we also constructed the transition pathways manually. Namely, 
\begin{itemize}
    \item The rotation path, which is defined by a change of the torsional angle $\phi$ around the central double bond (see Fig.~\ref{fig:Fig1});
    \item The inversion path, whose main feature is the bending of either $\theta$ or $\theta^\prime$ (see Fig.~\ref{fig:Fig1});
    \item The rotation assisted by inversion path, which is the combination of the first two.
\end{itemize}
Each path is comprised of 15 intermediate geometries linking the minima. In all cases, the molecule was forced to follow the desired mechanism by linearly interpolating the main degree(s) of freedom between both minima (tables S2-S4 show important geometric data and Fig. S1 the energy profiles). The obtained highest-energy geometries are in good agreement with those found elsewhere.\cite {Wang2007,Cembran2004,Gagliardi2004,Cattaneo1999}

\begin{table}[ht!]
\renewcommand{\arraystretch}{1.5}
    \centering
    \caption{\label{tab:Tab1}Relative energies ($\Delta$\textit{E}; in kcal mol\textsuperscript{-1}) of the highest-energy structures on each mechanism computed with the PBE+TS method.}
    \begin{ruledtabular}
    \begin{tabular}{cccccccc}
        Mechanism &  Rotation & Inversion & Rot+Inv  \\
        \hline
        $\Delta$\textit{E} & 30.2 & 27.4 & 27.5, 27.6 \\
    \end{tabular}
    \end{ruledtabular}
\end{table}

Table~\ref{tab:Tab1} shows the relative energies of the highest-energy structures found for each transition path. The rotation mechanism is the most favorable path at the initial and final steps of the isomerization, but it has the highest energy barrier among the three transition paths considered here. The inversion mechanism is the one with the lowest-lying highest-energy structure within PBE+TS calculations. The rotation assisted by inversion path is the least favored at the zones close to the minima and presents a plateau region at the top of the curve with two ``peaks'' with relative energies close to that of the highest-energy configuration in the inversion mechanism. 

The pathways introduced in the previous paragraphs are just linear interpolations between the {\itshape cis} and {\itshape trans} geometries. This introduces constraints on how the different degrees of freedom can evolve through the transition. To obtain a path affected by the contributions of all the important degrees of freedom, we ``optimized'' the rotation path by choosing the values of $\theta$, $\theta^\prime$, {\itshape a} and {\itshape b} that minimize the energy at each step. The optimized rotation path is the most favorable with the PBE+TS method with an energy barrier of 26.1 kcal mol\textsuperscript{-1}. Geometric details of this path can be found in table S5.

All the paths described here can be considered as good insights into the real isomerization process. It has been found that the activation barrier of the {\itshape cis} to {\itshape trans} thermal relaxation in n-heptane solution is between 22.7 and 25.1 kcal mol\textsuperscript{-1}.\cite {Bandara2012} So, in the gas phase we would expect greater values like those presented here. In what follows, we will focus on the optimized rotation (named simply rotation from now on) and the inversion mechanisms.

\subsubsection{Datasets.} 
Constructing a dataset for a molecule such as azobenzene requires a more elaborate procedure compared to the simpler glycine molecule. First of all, the transition process is a rare event at ambient conditions and cannot be easily accessed; second, there are more than one possible transition pathways.

\begin{figure}[!ht]
\includegraphics[height=.58\textheight]{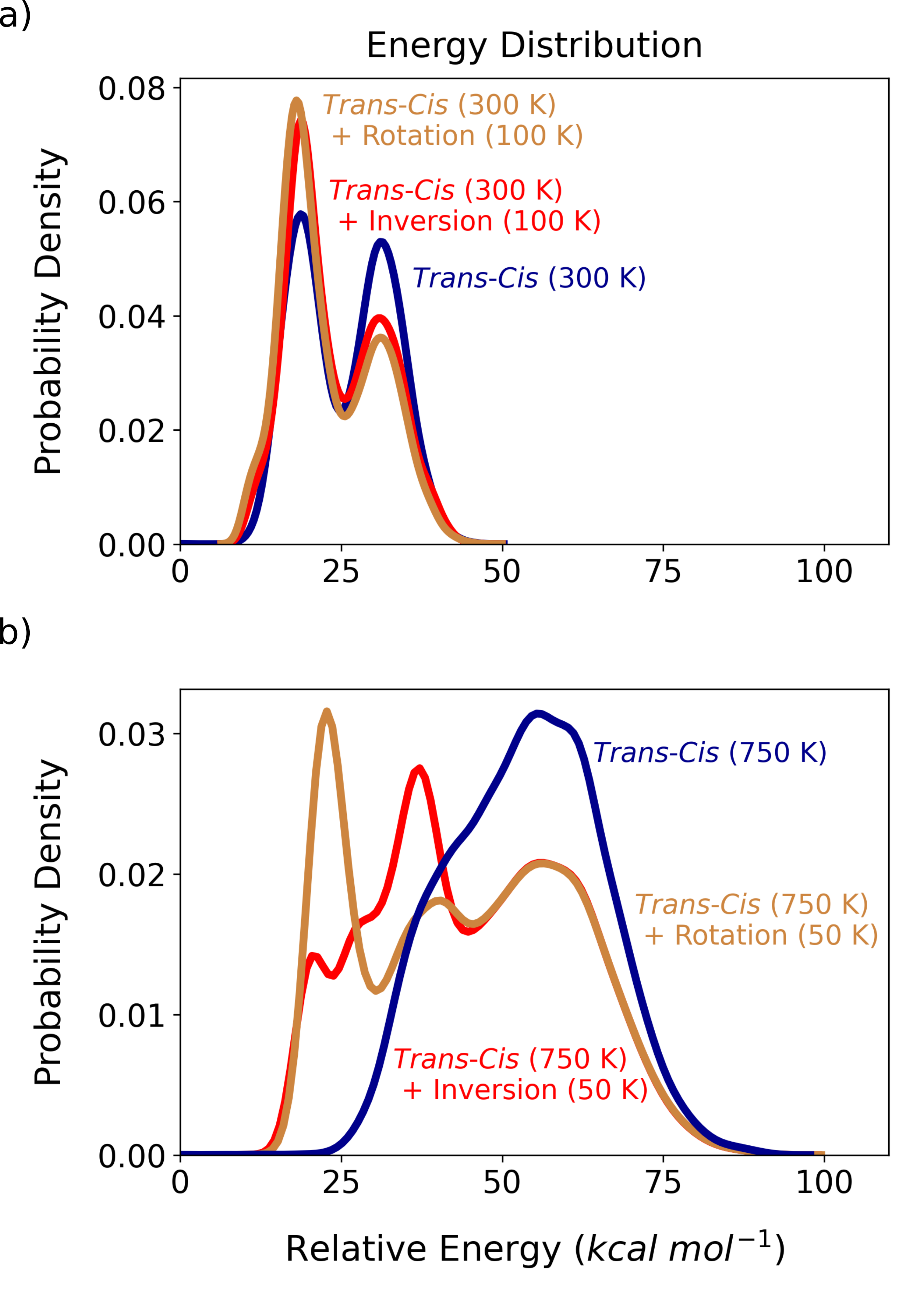}
\caption{\label{fig:Fig2} Distribution of relative energies (in kcal mol\textsuperscript{-1}; with respect to the optimized configuration of transazobenzene) of (a) ``close-to-equilibrium'' configurations at 300 K and geometries close to the transition paths at 100 K, and (b) high-energy ``close-to-equilibrium'' configurations and properly sampled geometries close to the transition paths.}
\end{figure}

Here, we build separate datasets for the rotation and inversion mechanism. We first combined two types of MD simulations: a) long constant-temperature MD runs with a time-step of 1~fs at 300~K at the PBE+TS/light level of theory starting from the equilibrium geometries, from which we selected a configuration every 25 steps and carried out single-point calculations with the PBE+TS/tight method (around 3500 configurations were collected for each minimum); b) constant-temperature MD runs of 300 steps with a 0.5~fs time-step at 100~K starting from each of the intermediate steps of the rotation and inversion paths. From the energy distributions shown in Fig.~\ref{fig:Fig2}a, one can conclude that the configurations visited during our MD simulations are bounded by the temperature to certain energy ranges, as indicated by the well-defined peaks representing \textit{trans} and \textit{cis}-like configurations. The addition of the configurations obtained from the rotation and inversion mechanism had little impact on the energy distribution, which lead us to conclude that the transition parts are still poorly sampled. Thus, we generated additional configurations by performing a) four constant-temperature MD simulations (of 2500 steps each) at 750~K with a time step of 1~fs starting from structures close to equilibrium and b) constant-temperature MD simulations of 2500 steps at 50~K with a very small time step (0.025~fs) starting from steps 7 and 9 of the inversion path, and the steps 8 and 10 of the rotation path (see tables S3 and S5 for details). The former provides the data required to model the cooling down process from transition states to minima, which involves high kinetic energies. The latter allows us to include slow changes of the degrees of freedom during the transition process. Fig.~\ref{fig:Fig2}b shows that the new reference geometries sample different energy distributions for close-to-equilibrium, rotation and inversion datasets. The final datasets combine the results of all 4 types of simulations containing 26455 data points for the rotation and 25528 data points for the inversion mechanisms.

An alternative to the above-described procedure would be to use an enhanced sampling technique (e.g. metadynamics, umbrella sampling). However, when using the dihedral $\phi$ and the angle $\theta$ as collective variables in a metadynamics simulation, we have observed that the trajectories explored from \textit{cis}- to \textit{trans}-like configurations (and vice-versa) do not have a clearly defined reaction coordinate. Hence, the datasets that we introduced in the previous paragraph, containing sampling of specific degrees of freedom, are more suitable to assess the performance of ML models on transition processes.

\section{Advantages and Limitations of Different ML Methods}\label{sec:section_III}
Before applying any ML method, the molecular configurations must be encoded into an appropriate rotationally, translationally and permutation invariant representation or ``descriptor''. There are many descriptors for MLFF available in the literature\cite{Bartok2015,Rupp2015,Eickenberg2018,Glielmo2017,Rupp2012,Bartok2010,Hansen2013,De2016,Artrith2017,Bartok2017,Yao2017,Faber2017,Glielmo2018,Grisafi2018,Tang2018,Pronobis2018,Faber2018,Christensen2020} and efforts to find suitable representations are still ongoing. We remark that a descriptor must balance efficiency with accuracy, hence different descriptors are applicable to different scenarios. One can divide them into local and global descriptors. For the former, neural networks (e.g. SchNet\cite{Schutt2017a,Schutt2018,Schutt2019}) and kernel-based potentials (e. g. when using descriptors such as FCHL\cite{Faber2018,Christensen2020} or SOAP\cite{Bartok2013}) assume locality through the introduction of a cutoff radius, and the interactions between atoms are modeled as a sum of individual atomic contributions. Conversely, global descriptors (like inverse pair-wise distances\cite{Chmiela2017,Rupp2012,Hansen2015}) can serve to build models where the prediction is obtained for the whole structure. Both approaches have their own advantages: for instance, while local descriptors can identify similar neighborhoods in small molecules that can be later transferred to larger systems, global descriptors can capture all interactions of a given system whenever the reference calculations contain the relevant information. However, descriptors also have their pitfalls, some of which arise with large flexible molecules and might become worse when dealing with complex processes, such as those happening along transition paths. 

\begin{figure}[!ht]
\includegraphics[height=.3\textheight]{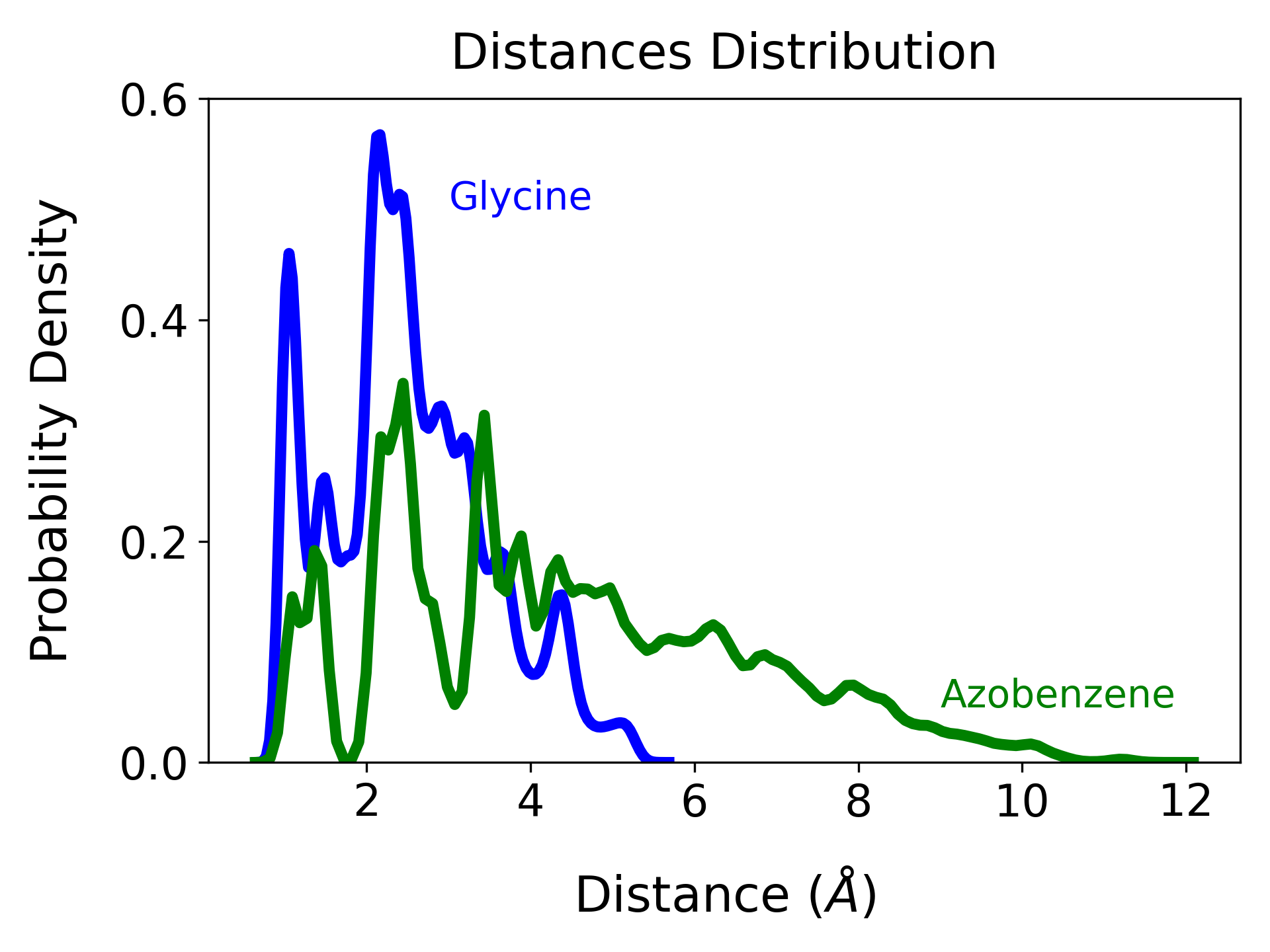}
\caption{\label{fig:Fig3} Distribution of interatomic distances (in \AA) in glycine and azobenzene rotation datasets.}
\end{figure}

The first immediate issue that one can foresee is the limited reach of local descriptors, as imposed by the selection of the cutoff radius. Fig.~\ref{fig:Fig3} shows the interatomic distance distribution in glycine and azobenzene rotation datasets. While for glycine the largest distances remain under 6 \AA\ and values lower than 4 \AA\ are the most populated, in azobenzene the distances present values of up to 12 \AA, and distances between 5 and 8 \AA\ are rather common. Thus, local descriptors might already face problems with molecules as large as azobenzene when relevant interactions fall outside their scope (see results with GAP/SOAP in section~\ref{sec:section_IV}). An example of such interactions are the long-range ones, which play an important role during azobenzene isomerization as suggested from the paths constructed in section~\ref{sec:section_II}. Specifically, vdW interactions decrease going from the {\itshape cis} to the {\itshape trans} configuration and these interactions lead to an increment of the energy barriers of all paths of more than 1.0 kcal mol\textsuperscript{-1} (details are shown in tables S2-S5). Increasing the cutoff radius appears to be a straightforward solution, but the potential gain in accuracy might lead to a significant loss in efficiency.

The second problem affects both local and global descriptors and is related to the scope of the descriptor itself. If some important features are not included or not properly represented in the selected descriptor, the method will fail to achieve appropriate performance (see results with sGDML in section~\ref{sec:section_V}). This could well apply to transition paths, where small changes in specific degrees of freedom of the molecule result in considerable energy variations. The addition of the relevant features to the descriptor might alleviate this issue, but requires \textit{a priori} knowledge of the studied system. 

\section{Accuracy of ML Models for Transition Paths}\label{sec:section_IV}
Although ML potentials have evolved successfully, there are many open challenges. Among others, the problem of building accurate and data-efficient ML models for flexible molecules describing equilibrium states and the transition pathways between them deserves special attention. Below we assess the performance of state-of-art ML models (BPNN,\cite{Behler2007,Behler2011} SchNet,\cite{Schutt2017a,Schutt2018,Schutt2019} GAP/SOAP\cite{Bartok2010,Bartok2015,Bartok2013} and sGDML\cite{Chmiela2017,Chmiela2018,Chmiela2019c,Sauceda2019}) on the PES of glycine and azobenzene molecules. The cutoff radius in BPNN, SchNet and GAP/SOAP was set to a typical value of 5 \AA, although additional tests with GAP/SOAP were done with larger cutoffs. All other important settings for each ML method are given in the supplementary material. Namely, we used (a) the glycine dataset, (b) the inversion and rotation datasets of azobenzene (see section~\ref{sec:section_II}). 

The training and test sets were created as follows: first, we used the training set selection process of sGDML (which is based on the Boltzmann distribution) to create subsets from each of the datasets used in this work. The subsets have a size equal to five times the number of training points (e.g. for 1000 training points a subset of 5000 configurations was constructed). We then performed 5-fold cross-validation on each subset, using a single fold for training and the rest for testing. For instance, when using 1000 training points, we tested our model with the remaining 4000 configurations (which represents around 15$\%$ of the datasets of azobenzene and 40$\%$ of the glycine dataset). We randomly created the cross-validation tasks while ensuring that in each fold the energy distribution of the whole dataset was preserved. In this way, each of the folds is representative of the original dataset, such that testing our models on these folds will give similar results as testing on the complete dataset. The proposed training/test set selection procedure is comparable to a default random scheme, as implemented in SchNet, while providing more reliable and accurate ML models.

\begin{figure}[!ht]
\includegraphics[width=.75\textwidth]{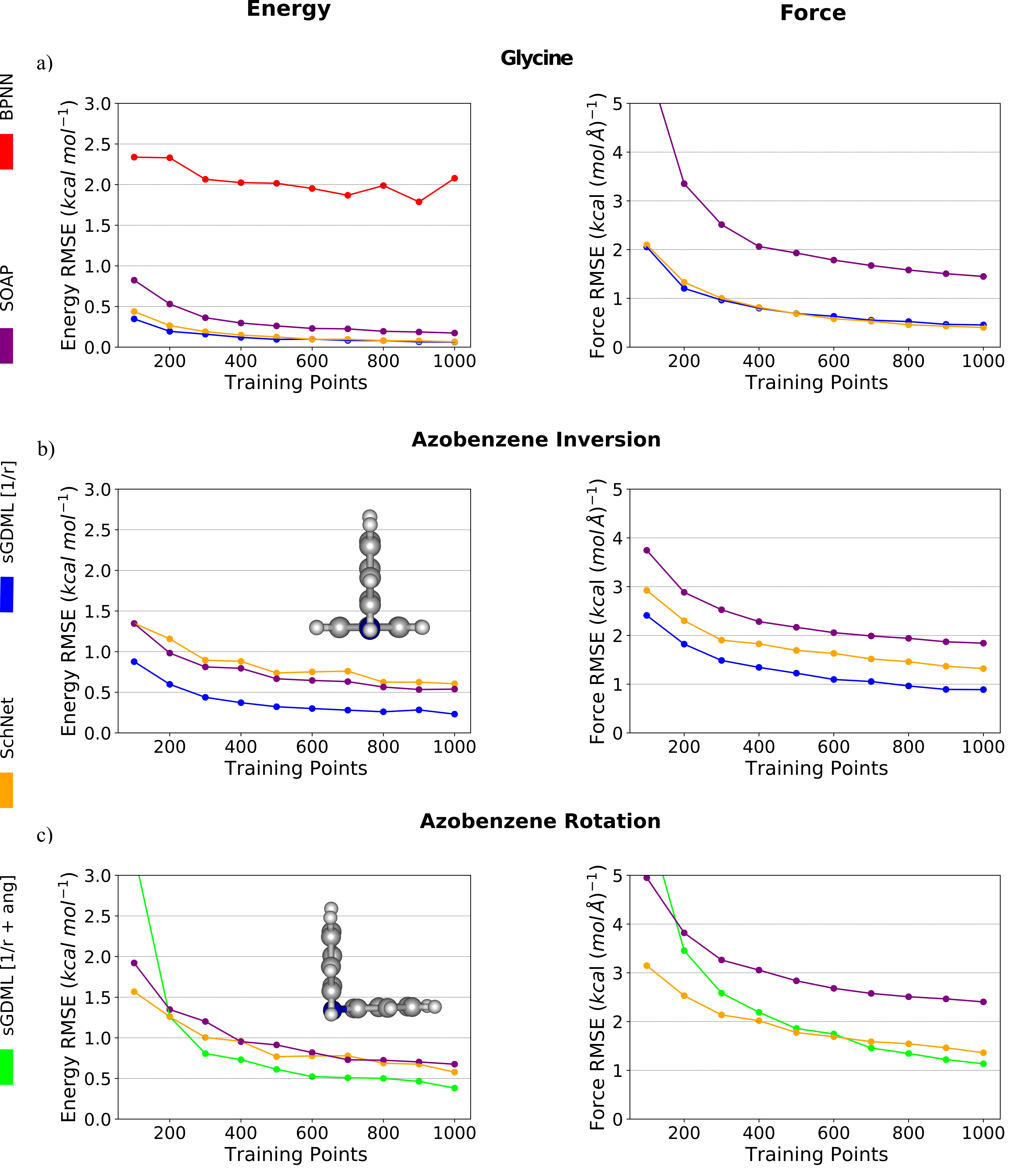}
\caption{\label{fig:Fig4} Energy (in kcal mol\textsuperscript{-1}) and force (in kcal (mol \AA)\textsuperscript{-1}) prediction accuracy in terms of root mean squared error (RMSE) as a function of training set size for a) glycine, b) inversion and c) rotation datasets of azobenzene using the best models of BPNN, GAP/SOAP, SchNet, and sGDML with the default descriptor (sGDML~[1/r]) or with the extended descriptor (sGDML~[1/r~+~ang]) out of all cross-validation tasks. Only models with errors below 5.0 kcal (mol \AA)\textsuperscript{-1} and 3.0 kcal mol\textsuperscript{-1} are shown. b) and c) show a side view of the highest energy structure on the inversion and rotation paths of azobenzene, respectively.}
\end{figure}

Fig.~\ref{fig:Fig4} shows the energy and force prediction accuracy for the best models out of all cross-validation tasks. For SchNet, BPNN and GAP/SOAP, the best model for a given training set size is the one showing the lowest energy root mean squared error (RMSE) in the test set. In the case of sGDML, the best model usually is the one with lowest force RMSE in the test set. However, if two or more sGDML models with similar force RMSE ($\sim$0.1 kcal (mol \AA)\textsuperscript{-1}) present substantially different energy RMSE ($\sim$1.0 kcal mol\textsuperscript{-1}), we favoured the one with the lowest energy RMSE. In the following paragraphs the errors we discuss correspond to those of the best models.

For the small glycine molecule (Fig.~\ref{fig:Fig4}a), BPNN presents high errors, with a RMSE above 6.0 kcal (mol \AA)\textsuperscript{-1} for forces and around 2.0 kcal mol\textsuperscript{-1} for energies, even after using 1000 training points. sGDML and SchNet perform much better with errors below 1.0 kcal mol\textsuperscript{-1} and 1.0 kcal (mol \AA)\textsuperscript{-1} with 300 and 400 training points, respectively. GAP/SOAP also shows a good performance in energy prediction with errors under 1.0 kcal mol\textsuperscript{-1} after using 100 training points, although it is less accurate when predicting forces (errors remain around 1.5 kcal (mol \AA)\textsuperscript{-1} with 1000 training points). Based on this analysis, we are henceforth not considering BPNN as a valid candidate to reproduce more complex PES of flexible molecules using a limited amount of training points, instead focusing on GAP/SOAP, sGDML, and SchNet.

For the azobenzene datasets the results for different models show high variability. GAP/SOAP obtains an error in energies under 1.0 kcal mol\textsuperscript{-1} for the inversion mechanism with only 200 training points (Fig.~\ref{fig:Fig4}b). However, for the rotation mechanism this performance is achieved with 400 training points (Fig.~\ref{fig:Fig4}c). Also, force prediction accuracy is worse for the rotation mechanism (remains above 2.4 kcal (mol \AA)\textsuperscript{-1} with 1000 training points) than for the inversion mechanism (remains above 1.8 kcal (mol \AA)\textsuperscript{-1} with 1000 training points) along the whole learning curves. This means that the parts of the PES that are covered by each transition process involve different contributions from the long-range interactions. Indeed, SOAP learns the local information (within the selected cutoff radius), which quickly saturates with the increase of the training set, but cannot capture relevant long-range interactions particularly important for the rotation mechanism.

sGDML achieves an outstanding performance for both transition mechanisms. For the inversion dataset, errors in energy go under 1.0~kcal mol\textsuperscript{-1} with 100 training points and errors in forces go under 1.0~kcal (mol \AA)\textsuperscript{-1} with 800 training points (Fig.~\ref{fig:Fig4}b). For the rotation dataset, 300 training points are needed to obtain an energy RMSE lower than 1.0~kcal mol\textsuperscript{-1} and the force RMSE gets close to 1.1~kcal (mol \AA)\textsuperscript{-1} after using 1000 training points. To achieve this performance, however, one requires different descriptors for different mechanisms: the default descriptor (inverse pair-wise distances) for the inversion mechanism and an extended descriptor (inverse pair-wise distances, and information on bonded angles and dihedrals in the form \(D\textsubscript{$\Theta$} = (1-e\textsuperscript{-$\Theta$})\textsuperscript{2} - 1\) and \(D\textsubscript{$\Phi$} = 1 + Cos\Phi\), respectively).  The features in a given descriptor are not evenly important for all parts of the PES. Thus, a single descriptor might be unable to properly resolve all relations between the relevant degrees of freedom for the PES and each of its transition mechanisms.

SchNet also achieves `chemical accuracy' of 1.0~kcal mol\textsuperscript{-1} for both inversion and rotation datasets but, contrary to sGDML, it can deal with both mechanisms using the same settings (Figs.~\ref{fig:Fig4}b,c). However, SchNet is less reliable than sGDML when predicting forces, with an overall RMSE of around 1.4 kcal (mol \AA)\textsuperscript{-1} for both the inversion and the rotation datasets with 1000 training points. Better performance can be expected with larger training sets. However, this is a trivial solution limited in practice by the increased computational costs of the reference data for larger molecular sizes. The reason SchNet outperforms GAP/SOAP is also clear: even though SchNet primarily learns local features, it can learn long-range interactions by embedding such features into the local environments for different parts of the molecule.

\section{Challenges for ML Models in Flexible Molecules}\label{sec:section_V}
Even though GAP/SOAP, sGDML and SchNet methods are able to learn the PES of the azobenzene molecule with chemical accuracy, there is a considerable difference between the predictions for different methods, as well as for different transition mechanisms within the same method. Below we demonstrate that these contrasting results are caused by imperfections of the implemented training set selection schemes (suboptimal for complex PES with multiple local minima), as well as intrinsic limitations of the employed descriptors (unable to equally capture short- and long-range interactions). To do so, we (i) explore the dependence of the performance of ML methods on the specific selection of training set by considering the results of all cross-validation tasks and (ii) modify the descriptors of GAP/SOAP and sGDML models.

\begin{figure}[!ht]
\includegraphics[width=.6\textheight]{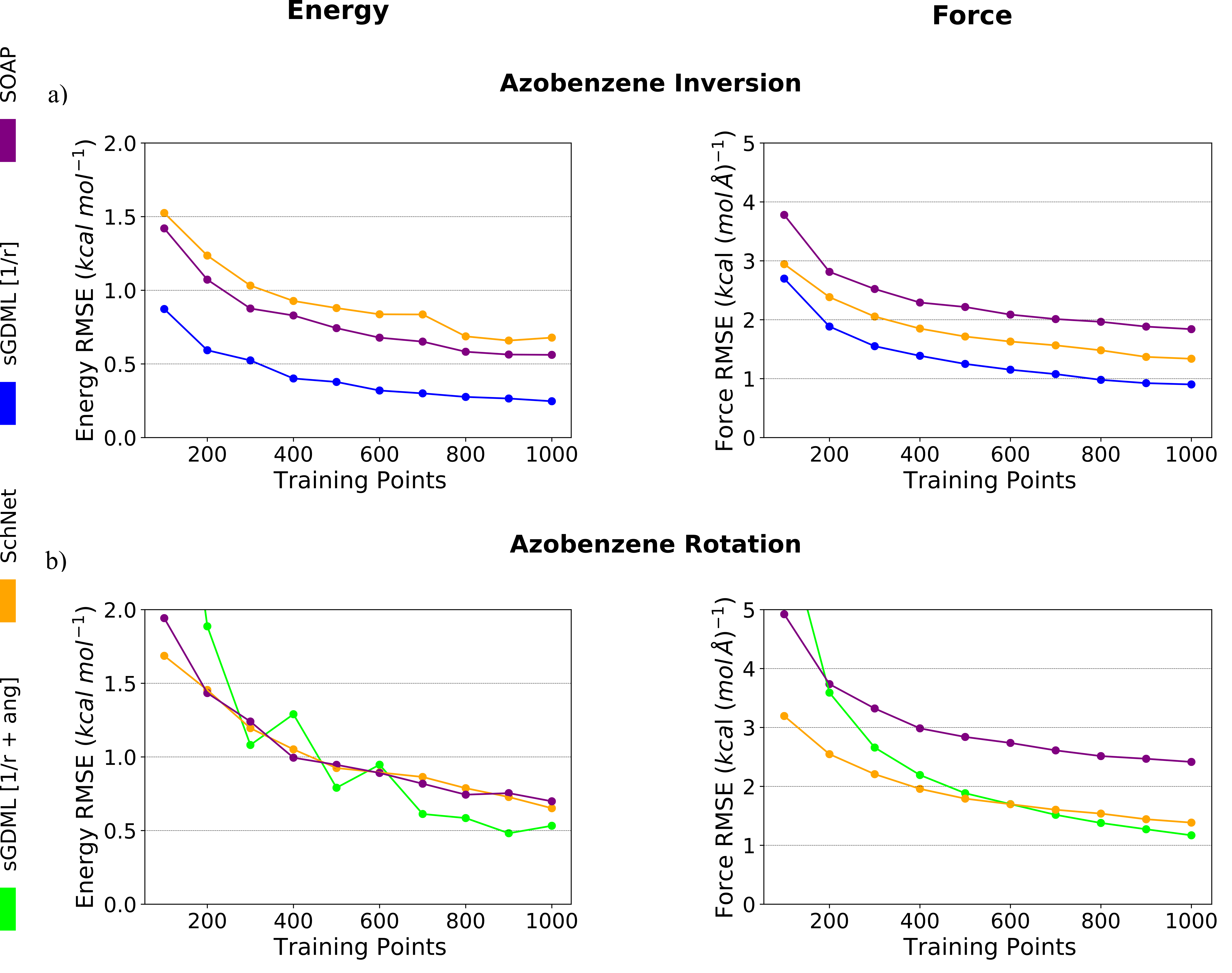}
\caption{\label{fig:Fig5} Mean of the energy (in kcal mol\textsuperscript{-1}) and the force (in kcal (mol \AA)\textsuperscript{-1}) prediction accuracy over all cross-validation tasks in terms of RMSE as a function of training set size for SchNet, SOAP, sGDML with the default descriptor (sGDML [1/r]) and with the extended descriptor (sGDML [1/r + ang]) for the inversion and the rotation datasets of azobenzene. Only models with errors below 5.0 kcal (mol \AA)\textsuperscript{-1} and 2.0 kcal mol\textsuperscript{-1} are shown. In the case of sGDML models we only show the curve of the best performing descriptor for each mechanism.}
\end{figure}

\begin{figure}[!ht]
\includegraphics[width=.6\textheight]{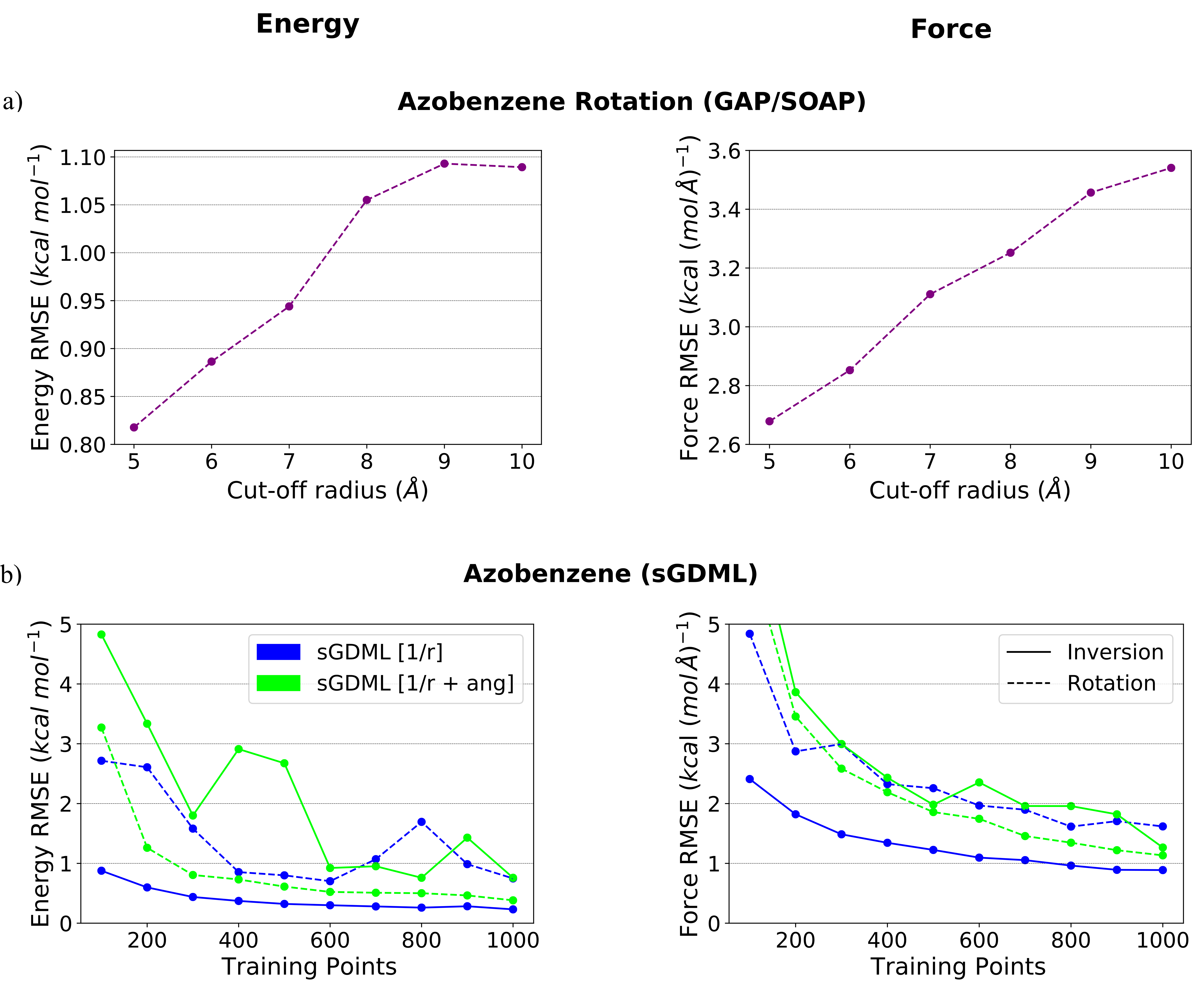}
\caption{\label{fig:Fig6} Energy (in kcal mol\textsuperscript{-1}) and force (in kcal (mol \AA)\textsuperscript{-1}) prediction accuracy in terms of root mean squared error (RMSE) a) as a function of cutoff radius for GAP/SOAP models of the rotation dataset of azobenzene trained with 600 training points; and b) as a function of training set size for sGDML with the default descriptor (sGDML [1/r]) and with the extended descriptor (sGDML [1/r + ang]) for the inversion and rotation datasets of azobenzene. Only models with errors below 5.0 kcal (mol \AA)\textsuperscript{-1} and 5.0 kcal mol\textsuperscript{-1} are shown.}
\end{figure}

The prediction accuracy of GAP/SOAP models is not considerably affected by the particular choice of training set of each cross-validation task. The average energy RMSE over all cross-validation tasks (see Fig.~\ref{fig:Fig5}) are practically the same as the RMSE of the best model (see Fig.~\ref{fig:Fig4}), with both errors going under 1.0~kcal mol\textsuperscript{-1} with 400 training points. The main shortcoming of GAP/SOAP models is in learning long-range interactions. This explains the $\sim$0.3 kcal mol\textsuperscript{-1} larger RMSE for the rotation mechanism (Fig.~\ref{fig:Fig5}). Indeed, due to the different mutual orientation of the benzene rings (see geometries of the highest energy structures on each transition path in Figs.~\ref{fig:Fig4}b,c), the vdW energy contribution along the rotation transition pathway is, in average, larger by $\sim$0.3 kcal mol\textsuperscript{-1} than that for the inversion one. Increasing the cutoff radius while keeping the same number of basis functions does not resolve the challenge. Fig.~\ref{fig:Fig6}a shows the best energy and force RMSE as a function of the cutoff radius in GAP/SOAP models with 12 radial and 6 angular functions. In fact, both energy and force prediction become slightly worse. Increasing the number of basis functions to handle this issue would lead to computationally expensive ML models, impractical for realistic applications. As an alternative approach to include long-range interactions one can use the recently developed LODE method,\cite{LODE} but such calculations are out of the scope of this work.

The sGDML model provides the most accurate and data-efficient FF, but faces two important issues. The first one is the descriptor, as shown by Fig.~\ref{fig:Fig4}, where we had to employ different descriptors for different transition mechanisms. Specifically, for the inversion mechanism the default sGDML descriptor (inverse pair-wise distances) is sufficient. In contrast, a reliable description of the rotation mechanism requires the inclusion of information about angles and dihedrals in the form \(D\textsubscript{$\Theta$} = (1-e\textsuperscript{-$\Theta$})\textsuperscript{2} - 1\) and \(D\textsubscript{$\Phi$} = 1 + Cos\Phi\), where $\Theta$ and $\Phi$ are any bonded angle and dihedral of the molecule in radians respectively. The performance of the default and extended descriptors is shown in Fig.~\ref{fig:Fig6}b. The default descriptor for the rotation mechanism and the extended descriptor for the inversion mechanism present considerable oscillations in the energy error as a function of training set size (Fig.~\ref{fig:Fig6}b), which is unacceptable behavior for a reliable ML model. The reason is simple: on one hand, inverse pair-wise distances cannot correctly resolve the states along the transition path of the rotation mechanism, which are defined by changes in the dihedral angle $\phi$. On the other hand, all angles and dihedrals are not equally representative of the inversion mechanism and adding them misleads the model in this case. 

We would like to remark that even when using an appropriate descriptor, the training set selection is still crucial. Out of all the methods, sGDML is the most affected by the model selection in cross-validation tasks (compare Fig.~\ref{fig:Fig4}c and Fig.~\ref{fig:Fig5}b). While for 1000 training points the difference between the average energy RMSE over all cross-validation tasks and the energy RMSE of the best model is of only 0.2~kcal mol\textsuperscript{-1}, for 400 training points this difference is as large as 0.6~kcal mol\textsuperscript{-1}. Thereby one needs to be very careful when selecting the best sGDML models. Furthermore, the training set dependency often leads to models with similar force RMSE but considerably different energy RMSE. For instance, the average difference in energy RMSEs between the best and the worst models for the rotation mechanism using the extended descriptor (over all training set sizes considered here) is of 0.7~kcal mol\textsuperscript{-1}, while the average difference in force RMSEs is less than 0.2~kcal (mol \AA)\textsuperscript{-1}. To understand this behaviour, one need to recall that sGDML models contain two hyperparameters apart from the regularization. One is the width of the kernel, which is defined by optimizing the force predictions. The other one is the constant shift for the energy, which is employed to minimize the difference between the prediction results and the energy values in the dataset. The energies of flexible molecules are highly degenerated. Consequently, even though all our training sets follow the energy distribution of the complete dataset, they represent different parts of the PES unequally. As a result, the energy shift hyperparameter obtained from a given training set can be far from optimal for the whole dataset. Hence, the force-based best model selection scheme, as implemented in sGDML, may lead to large oscillation in energy prediction accuracy as a function of training set size (similar to those in the green solid line in Fig.~\ref{fig:Fig6}b). To resolve this issue, one should consider both energy and forces to select the optimal model. For each cross-validation task, the training scheme does not change and still relies only on forces, but for selecting the best model out of all, we also account for the energy prediction accuracy. Summarizing, accurate and data-efficient models are achievable with sGDML, but both descriptors and training sets must be carefully selected.

\begin{figure}[!ht]
\includegraphics[width=.75\textwidth]{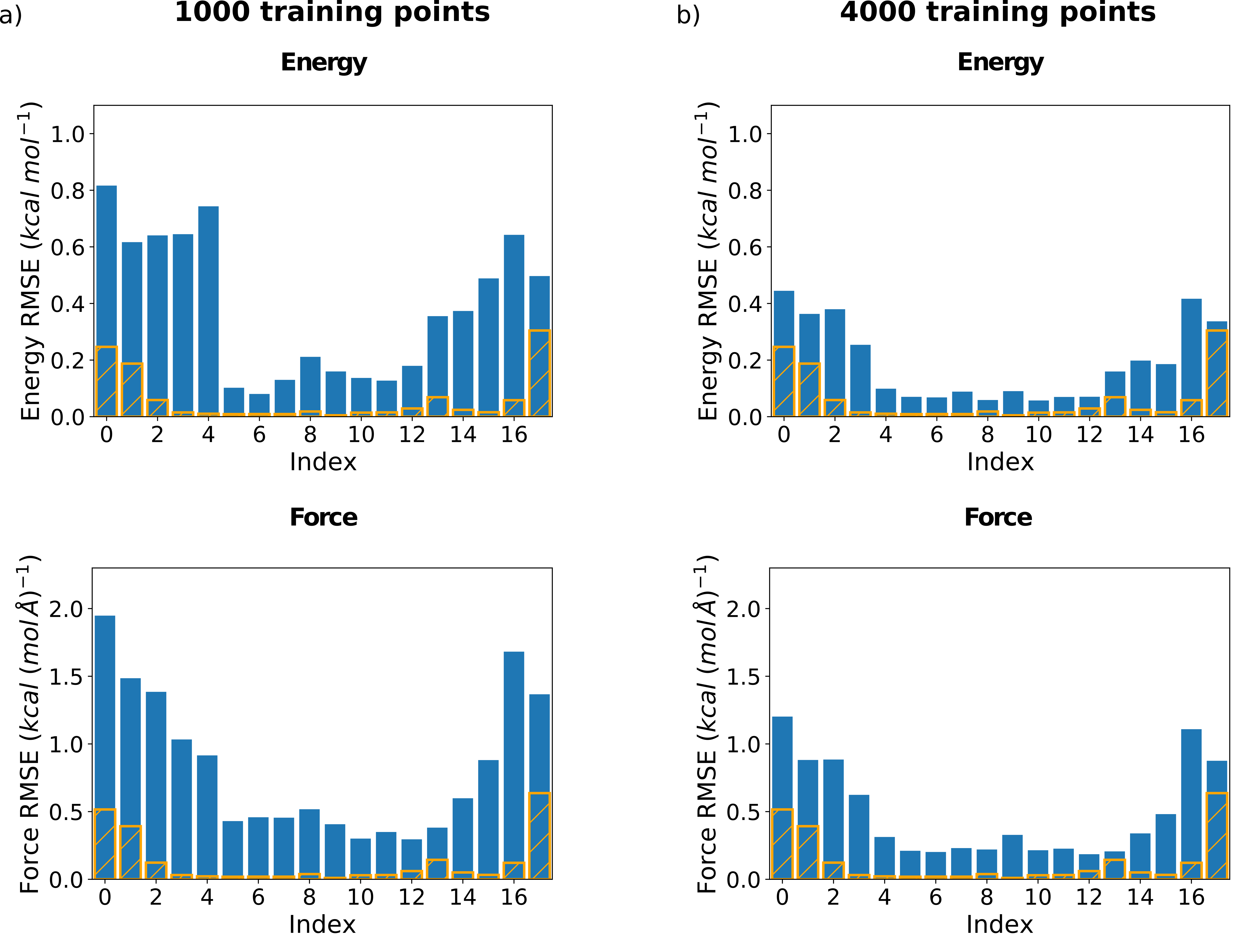}
\caption{\label{fig:Fig7} Energy (in kcal mol\textsuperscript{-1}) and force (in kcal (mol \AA)\textsuperscript{-1}) prediction accuracy in terms of root mean squared error (RMSE) for different clusters of the rotation dataset using a) the best SchNet model obtained with 1000 training points of the rotation dataset (RMSE of 0.6 kcal mol\textsuperscript{-1} and 1.3 kcal (mol \AA)\textsuperscript{-1} over the complete rotation dataset) and b) a SchNet model trained with 4000 configurations combining the rotation and inversion datasets (RMSE of 0.4 kcal mol\textsuperscript{-1} and 0.8 kcal (mol \AA)\textsuperscript{-1} over the complete rotation dataset). The configurations for each cluster were selected depending on the value of the dihedral angle $\phi$, going from an interval between 0$^{\circ}$ and 10$^{\circ}$ for index 0, to an interval between 170$^{\circ}$ and 180$^{\circ}$ for index 17. Relative population of each cluster is also indicated (solid orange lines, arbitrary units).}
\end{figure}

SchNet is an optimal compromise between GAP/SOAP and sGDML models. Like GAP/SOAP, it does not heavily depend on the specific selection of training set (see Fig.~\ref{fig:Fig5}), while being capable to learn long-range interactions, akin to sGDML (see Fig.~\ref{fig:Fig4}). As a result, SchNet reproduces both transition mechanisms equally accurately using the same settings, with errors only slightly larger than those of the sGDML models. This is a consequence of the embedding of local features of different atoms through the interaction layers. To train our SchNet models, we employed 6 interaction layers with a 5 \AA\ cutoff radius for local environment. This architecture guarantees that we cover all possible interatomic distances within azobenzene molecule (Fig.~\ref{fig:Fig3}), making our SchNet models effectively global. However, a good overall RMSE might not always mean a good ML model. Fig.~\ref{fig:Fig7}a shows the energy and force prediction accuracy on different clusters of the rotation dataset of the best SchNet model out of the 5 cross-validation tasks with 1000 training points (RMSE of 0.6 kcal mol\textsuperscript{-1} and 1.3 kcal (mol \AA)\textsuperscript{-1} over the complete rotation dataset). Each cluster corresponds to different values of the dihedral angle $\phi$ (from the interval between 0$^{\circ}$ and 10$^{\circ}$ for index 0 to the interval between 170$^{\circ}$ and 180$^{\circ}$ for index 17). One can see that the errors for close-to-equilibrium configurations are four times larger than those for the transition regions. Increasing the training set size and having information of the two mechanisms (i.e. we added the data of the inversion mechanism to the rotation dataset) does not change this ratio (Fig.~\ref{fig:Fig7}b). We remark that while building these models, we expect to be able to use them for accurate simulation of the transition mechanism computing reaction rates, lifetime of \textit{cis} configuration, etc. To do that, one needs a MLFF equally accurate for all relevant parts of the PES. As one can see from Fig.~\ref{fig:Fig7}, this requirement is not fulfilled by the obtained SchNet FFs. Importantly, by using a total of only 600 training points for both \textit{trans}- and \textit{cis}-like configurations (300 for each isomer), one can train a single sGDML model that reproduces the performance of the 4k SchNet model on clusters 0,1,2,16 and 17. Hence, while learning the entire PES of the azobenzene molecule is possible within a SchNet model, this approach is not data-efficient. There are two options to solve this problem: one is to use a training set optimization technique, flattening the prediction across the configuration space.\cite{Gregory} The second option would be to design schemes that combine a set of local models into a global one, finding optimal descriptors, training sets, and models for each part of the PES.

Summarizing the results of this section, while state-of-the-art ML models are capable of reproducing the complex PES of flexible molecules, this challenge is far from being solved in practice. Default approaches demonstrating excellent performance for small molecules or rigid systems struggle with increasing flexibility and dimensionality. Even the best performing models present difficulties to efficiently learn the PES in its entirety. To overcome this challenge, one should ensure that the descriptors contain all relevant features to capture the complex geometrical transformations in the high dimensional PES. Also, training sets must represent all parts of configuration space, which exhibits high energetic degeneracy, which makes purely following an energy distribution ineffective. Thus, further developments of robust approaches for selecting training points, appropriate descriptors, or even using different models for different parts of the PES are required.

\section{Conclusions}\label{sec:conclusions}
In the present work, we discussed the challenge of modeling the PES of flexible molecules using state-of-the-art ML models when using limited sets of training data. Our results show that methods based on local descriptors (e.g. BPNN and GAP/SOAP) saturate quickly with the increase in the number of training points, while not achieving the desired prediction accuracy. This is a consequence of their intrinsic limitations in describing long-range interactions. The ML methods based on global descriptors (e.g. sGDML) suffer in performance when the reference datasets consist of several disconnected parts of the PES. The main reasons are the inability of the default training schemes to select appropriate training datasets in unbalanced reference data and the limitations of the standard molecular descriptors to pick up the features describing the complex geometric transformations in flexible molecules. Finally, end-to-end NNs (e.g. SchNet) struggle to reproduce all relevant parts of the PES with equal accuracy. Moreover, NNs require overfeeding the models with training data, which results in unreasonable computational costs to generate expensive reference datasets. 

All of the tested MLFF in their current form can be further improved for quantitative studies of complex processes in flexible molecules. Important features of ML, such as descriptors and selection of training points, should be revised. The different behavior of most of the models for the rotation and inversion mechanisms in azobenzene also suggests switching from learning the entire PES within a single task to the employment of multiple local models for different parts of the PES with further combining them into a global FF.

\begin{acknowledgments}
VVG acknowledges financial support from the Luxembourg National Research Fund (FNR) under the program DTU PRIDE MASSENA (PRIDE/15/10935404). GF acknowledges financial support from the FNR AFR project 14593813. AT received support from the European Research Council (ERC-CoG BeStMo).
\end{acknowledgments}

\bibliography{References}

\end{document}